\documentclass{aa}  

\usepackage{graphicx}
\usepackage{txfonts}
\usepackage{amsmath}
\usepackage{xcolor}
\def\EXO {EXO\,0748--676}
\begin{document}

   \title{An ultraviolet burst oscillation candidate from the low-mass X-ray binary \EXO}


   \author{A. Miraval Zanon
          \inst{1}
          \and
          F. Ambrosino \inst{2}
          \and
          G. Illiano \inst{3,2}
          \and
          A. Papitto \inst{2}
          \and
          G.L. Israel \inst{2}
          \and
          F. Coti Zelati \inst{4,5,3}
          \and
          L. Stella \inst{2}
          \and
          T. Di Salvo \inst{6}
          \and
          S.\,Campana \inst{3}
          \and
            G. Benevento \inst{7}
          \and
          N.O. Pinciroli Vago \inst{8,2}
          \and
            M.C. Baglio \inst{3}
        \and
            P. Casella \inst{2}
        \and
            P. D'Avanzo \inst{3}
        \and
            D.\,de\,Martino \inst{9}
        \and
            M.\,Imbrogno \inst{2}
        \and
            R. La Placa \inst{2}
        \and
            S.E. Motta \inst{3,10}
          }

   \institute{
        ASI - Agenzia Spaziale Italiana, Via del Politecnico snc, 00133 Roma, Italy\\
              \email{arianna.miraval@asi.it}
         \and
             INAF, Osservatorio Astronomico di Roma, Via Frascati 33, 00078 Monteporzio Catone, RM, Italy  
        \and
             INAF, Osservatorio Astronomico di Brera, Via E. Bianchi 46, 23807 Merate, LC, Italy
        \and
            Institute of Space Sciences (ICE, CSIC), Campus UAB, Carrer de Can Magrans s/n, 08193 Barcelona, Spain
        \and
            Institut d’Estudis Espacials de Catalunya (IEEC), 08860 Castelldefels (Barcelona), Spain
        \and
            Universit\'a degli Studi di Palermo, Dipartimento di Fisica e Chimica, Via Archirafi 36, 90123 Palermo, Italy
        \and
            INFN, Sezione di Roma 2, Università di Roma Tor Vergata, Via della Ricerca Scientifica 1, 00133 Roma, Italy
        \and
            Department of Electronics Information and Bioengineering, Politecnico di Milano, 20133 Milano, Italy
        \and
            INAF, Osservatorio Astronomico di Capodimonte, Salita Moiariello 16, 80131 Napoli, Italy
        \and
            University of Oxford, Department of Physics, Astrophysics, Denys Wilkinson Building, Keble Road, OX1 3RH, Oxford, United Kingdom
             }

   \date{}

 
  \abstract
{X-ray burst oscillations are quasi-coherent periodic signals at frequencies close to the neutron star spin frequency.  They are observed during thermonuclear Type I X-ray bursts from a number of low-mass X-ray binaries (LMXBs) hosting a fast-spinning, weakly magnetic neutron star. Besides measuring the spin frequencies, burst oscillations hold the potential to accurately measure neutron star mass and radius, thus providing constraints on the equation of state of matter at nuclear densities. Based on far-ultraviolet (FUV) observations of the X-ray binary \EXO\ taken with the {\it Hubble Space Telescope} in 2003, we report a possible indication of ultraviolet burst oscillations at the neutron star spin frequency ($\sim$552\,Hz), potentially the first such case for an LMXB. The candidate signal is observed during an $\sim$8\,s interval in the rising phase of an FUV burst, which occurred $\sim$4\,s after a Type I X-ray burst. Through simulations, we estimated that the probability of detecting the observed signal power from pure random noise is $3.7\%$, decreasing to $0.3\%$ if only the burst rise interval is considered, during which X-ray burst oscillations had already been observed in this source. The background-subtracted folded pulse profile of the candidate FUV oscillations in the (120--160\,nm) band
is nearly sinusoidal with a $\sim 16$\% pulsed fraction, corresponding to a pulsed luminosity of $\sim 8 \times 10^{33}$\,erg\,s$^{-1}$. Interpreting the properties of this candidate FUV burst oscillations in the light of current 
models for optical-ultraviolet emission from neutron star LMXBs faces severe problems. If signals of this kind are confirmed in future observations, they might point to an unknown 
coherent emission process as the origin of the FUV burst oscillations observed in \EXO.
}


   \maketitle
%

\section{Introduction}

  Most low-mass X-ray binaries (LMXBs) host an old, weakly magnetic  ($\sim 10^8$\,G) neutron star (NS) which accretes matter from a late-type companion (<1\,M$_{\odot}$). NSs rapidly rotating on millisecond periods, as the result of spin-up by matter accretion torques over  
$\sim 10^8-10^9$~yr-long evolutionary timescales, are believed to be the progenitors of radio millisecond pulsars \citep[MSPs;][]{Alpar1982}. Coherent millisecond X-ray pulsations at the NS spin period have been found during the outbursts of about $\sim 25$ LMXBs \citep{DiSalvo2022}. Similar to longer-period, high-magnetic field X-ray pulsars, these pulsations result from funnelled accretion of matter onto the NS magnetic poles and testify to the presence of a magnetosphere. Type I X-ray bursts,  a characteristic feature of many NS LMXBs, are bright events with typical durations of tens of seconds,  blackbody-like spectra with typical temperatures of $\sim 1 -3$ keV and peak X-ray luminosity of $\sim 10^{38}$\,erg\,s$^{-1}$, close to or even reaching the Eddington limit of the NS. 
They are the result of thermonuclear flashes, where newly accreted material on the surface of the NS suddenly burns \cite[see, e.g., ][]{galloway21}, with characteristic recurrence times of a few hours. 
Transient quasi-coherent oscillations in the millisecond range have been observed during Type I X-ray bursts in a number of LMXBs \citep[see, e.g., the review by][]{Bhattacharyya2022}. These {\it Burst Oscillations} (hereafter BOs) are linked to the NS rotation as demonstrated by the near coincidence (within $\sim 1$\%) between their period and that of the coherent pulsations of magnetospheric origin that are present in the persistent emission of burst sources \citep{Strohmayer1996, chakrabarty2003}. BOs thus provide a tool to measure the NS spin frequency of LMXBs that do not display coherent pulsations.

In the most common interpretation, BOs are generated by a hot spot on the NS surface during a thermonuclear flash, with the observed flux oscillations being produced by the very rapid rotation of the NS through the combined effects of relativistic Doppler-like modulation, beaming, gravitational light deflection and possibly self-eclipsing \citep{poutanen2006, Morsink2007}.
Frequency drifts may result from the hot spot shifting in longitude (and perhaps in latitude), similar to cyclones on Earth \citep{Strohmayer1997, Spitkovsky, Watts2012, Chambers2020}. Spectrally resolved BO observations with a high signal-to-noise ratio using the next generation of very large area X-ray instruments are expected to provide high precision measurements of NS masses and radii, thus further constraining the equation of state of matter at supranuclear densities \citep{watts2016, Piro2022, Kini}.

The transient NS LMXB \EXO\;was discovered during an outburst in 1985 with the {\it European X-ray Observatory SATellite}, {\it EXOSAT}  \citep{Parmar1985}. The source remained continuously active until August 2008 \citep{Wolff2008, Hynes2008, Torres2008, Degenaar2011}, when it switched to quiescence. The source turned again into outburst in June 2024 \citep{Baglio2024_1, Knight2025}.
Observations while in outburst led to the detection of a number of Type I X-ray bursts \citep{Gottwald1986, Parmar1986}, many of which were studied with the {\it Rossi X-ray Timing Explorer} \citep[\textit{RXTE};][]{Bilous2019}.
In February 2003, a Type I burst of \EXO\; was observed simultaneously in the X-ray, far ultraviolet (FUV), and optical bands \citep{Hynes2006}. The similarity of the light curve profiles in the different bands, together with the $\sim$ 4\,s delay of the optical/FUV burst relative to the X-ray burst, suggested that the latter drove the burst emission in the lower energy bands. 
The tail of the burst exhibited a longer decay in the FUV and optical bands than in the X-rays; the rising phase was instead similar across the different energy bands. 
A very similar phenomenology has been observed also in other NS LMXBs and the optical/UV burst was interpreted in terms of reprocessing of the X-ray burst from the disk and/or the companion star \citep{Grindlay1978, McClintock1979, Hackwell1979, Koyama1981, Pedersen1982, Lawrence1983,Robinson1997, Hynes2006, Pearson2006}.

In 2007, X-ray BOs in two bursts of \EXO\;were discovered with {\it RXTE} around a frequency of $\sim$ 552\,Hz corresponding to a 
NS spin period of $\sim$ 1.8\,ms \citep{Galloway2010}. The peak-to-peak amplitude was $\sim 21\%$, and over the few seconds during which the oscillations were detected, the frequency drifted toward higher values by up to 2\,Hz. 
Searches for coherent pulsations in the persistent X-ray source flux failed to detect a significant signal at the NS spin frequency \citep{Jain2011}. 

\EXO\;also exhibits deep X-ray eclipses and partial optical eclipses every 3.82\,hr \citep{Crampton1986, Parmar1986}. The $\sim 500$ s-long X-ray eclipses are sharp and result from the companion star passing in front of the X-ray source \citep{Wolff2009}. They display a residual X-ray flux of $\sim 7-10$\% of
the corresponding mean flux outside eclipse, likely due to 
photons being scattered along the line of sight by an extended accretion disk corona \citep{Parmar1986, Bonnet-Bidaud2001}. The system has a high inclination, estimated to be approximately $77^{\circ}$ 
\citep{Knight2022}.  
 Like other LMXBs observed from a high inclination, \EXO\; also shows X-ray flux dips centered around orbital phases of $\sim 0.6$ and $\sim 0.9$, whose profile varies from one orbit to another \citep{Parmar1986}. These dips probably result from X-ray absorption by irregular bulging regions close to the impact point of the accretion stream from the companion star and the outer disk rim \citep{Homan2003}.

 In this paper, we discuss a possible candidate for a FUV burst oscillation found during a Type I X-ray burst from \EXO, which may represent the first hint of such a phenomenon in a Type I bursting source. In Sect. \ref{Obs} we describe the February 2003 {\it Hubble Space Telescope} ({\it HST}) FUV and {\it RXTE} X-ray data that we used and their reduction.
Section \ref{Timing} is devoted to the timing analysis
of the burst as observed with both facilities. 
In Sect.\,\ref{Discussion} we discuss some interpretations of our results. We draw our conclusions in Sect. \ref{Conclusion}.

\section{Observations}
\label{Obs}

\subsection{Hubble Space Telescope}

We analyzed archival {\it HST} observations of \EXO\ acquired on February 18--19, 2003 (proposal ID 9398) with the Space Telescope Imaging Spectrograph (STIS, \citealt{STIS}). These observations were performed in TIME-TAG mode using the FUV-MAMA detector with 125\,\textmu s time resolution. Data were acquired with the G140L grating (115--173\,nm) equipped with a 52 $\times$ 0.5\,arcsec slit \citep{Bohlin2001}. On-source data were collected during six consecutive visits for a total exposure time of about 7.5\,hr. In the fourth visit, the first and only UV burst in \EXO\ was detected \citep{Hynes2006}.
We employed the \texttt{stis$\_$photons} package\footnote{\url{https://github.com/Alymantara/stis_photons}} to correct the position of slit channels and assign the correct wavelengths to each time of arrival (ToA). We selected ToAs belonging to channels 993--1007 of the slit to isolate the source signal and minimize the background contribution, and in the 120–160\,nm wavelength interval to avoid noisy contribution due to the poor response of the G140L grating at the edge wavelengths. We then applied barycentric correction using the \texttt{ODELAYTIME} task (subroutine available in the IRAF/STSDAS software package) by employing the source coordinates R.A. (J2000)=$07^{\mathrm{h}}48^{\mathrm{m}}33\fs73$, DEC. (J2000)=$-67\degr45'07\farcs9$ \citep{Torres2008} and the JPL-DE200 ephemeris.
We extracted the source energy spectrum during 
the burst and reduced it with \texttt{INTTAG} and \texttt{CALSTIS} pipeline software, and then corrected for the interstellar extinction assuming E(B-V) = 0.06 \citep{Hynes2006}. 

\subsection{\textit{Rossi X-ray Timing Explorer}}

We analyzed the {\it RXTE} observation of \EXO\; that started on 2003 February 19 at 01:47\,UT (Obs ID 70047-02-01-11) and was carried out simultaneously with {\it HST} observations.
We reduced data acquired with the Proportional Counter Array (PCA; 2-40 keV; \citealt{Jahoda1996}) and 122\,\textmu s time resolution. 
Photon arrival times were corrected to the Solar System barycenter by using the \texttt{faxbary} tool along with the JPL-DE200 ephemeris and the source coordinates given above. 
The 3--20\,keV light curve of the X-ray burst is shown in Fig.\,\ref{Dyn_pow} in black. We extracted the spectrum of the source X-ray emission during the thermonuclear burst.
We considered data taken by all layers of the proportional counter units (PCU) 0-2-3, producing a response matrix with the latest issued calibration. We rebinned the spectrum with the FTOOL \texttt{grppha} in order to have at least 25 counts/bin and added a systematic error of 0.5\% to each spectral bin as suggested by the {\it RXTE} calibration team\footnote{\url{https://heasarc.gsfc.nasa.gov/docs/xte/XTE.html}}. The background was subtracted using the Sky$\_$VLE model (pca$\_$bkgd$\_$cmbrightvle$\_$eMv20051128.mdl).

\section{Timing analysis}
\label{Timing}
\subsection{Burst timing analysis}
\label{Analysis}

To search for BOs in the {\it HST} burst we broadly followed the procedure that \cite{Bilous2019} employed to estimate the significance of BOs in the {\it RXTE} dataset.
We computed Leahy-normalized power density spectra (PDS) using overlapping sliding windows of $t_{win}=8\,$s of length, stepped by 1\,s and covering the total burst duration of 127\,s. The burst duration was estimated by measuring both the mean ($\mu$) and standard deviation ($\sigma$) of the light curve distribution of pre-burst count rates binned at 1-s. The end of the burst was identified as the point in the burst tail where the count rate is just below $\mu$ + 2$\sigma$. The selected time interval of the burst used for the analysis was 52689.07727--52689.07874\,MJD. Since X-ray BOs of \EXO\; were previously detected around 552\,Hz and displayed a frequency drift\footnote{Orbital Doppler frequency shifts are negligible compared to the observed drift.} $<2$\,Hz, we conservatively searched for FUV oscillations over a frequency range
of $550 - 555$~Hz. Considering the PDS Fourier resolution of $1/t_{win}=1/8$ Hz, $N_{\nu}=40$ frequency trials were made in each PSD to cover such a frequency range. To identify potential BO candidates, we thus considered signals above a threshold power $P_{th}$, corresponding to a probability of getting $p_{th}=5\times10^{-4}$ chance candidates per single PSD. Assuming that in the relevant frequency range, variability in the absence of signal can be described by pure Poisson noise, the threshold power would be $P_{th}=2 \ln{N_{\nu}} + 2\ln({1/(1-C)}) = 22.6$, where $C$ represents the confidence level, set at 99.95\% \citep{Vaughan1994}.
 
\begin{figure}
   \centering
   \resizebox{\hsize}{!}{\includegraphics{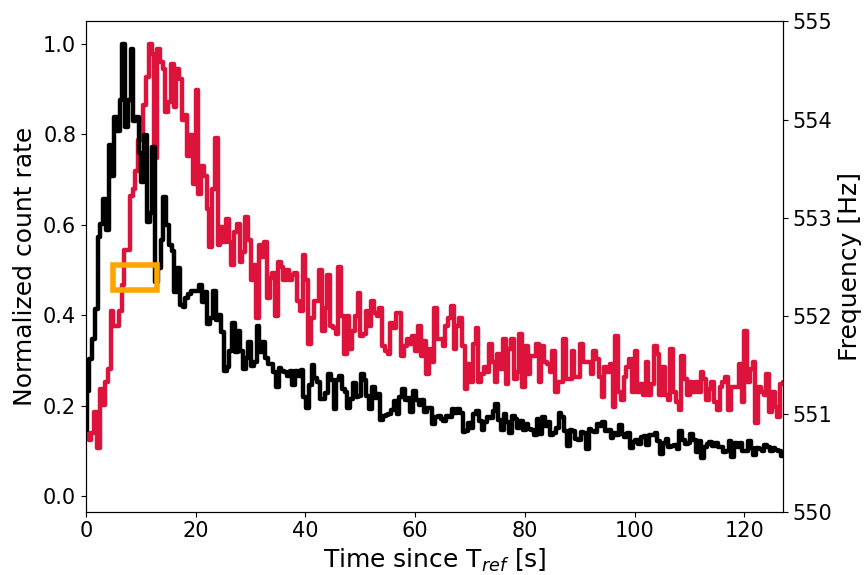}}
   \centering
      \caption{The red and black lines represent the FUV burst light curve observed by {\it HST} and the X-ray burst light curve detected by {\it RXTE}, respectively. The burst count rate is normalized for each instrument at the maximum count rate (312\,counts s$^{-1}$ for {\it HST} and 2949\,counts s$^{-1}$ for {\it RXTE}). The reference epoch is $T_{ref} = 52689.0772718$\,MJD. 
      The yellow box indicates the time interval of the BO candidate. The box size corresponds to the duration of the time window and the frequency resolution.} 
         \label{Dyn_pow}
   \end{figure}

During the final part of the burst rise (marked by the yellow box in Fig.~\ref{Dyn_pow}), a peak in the PDS was observed exceeding \(P_{\mathrm{th}}\), with a power of 23.17 at a frequency of 552.39(6)\,Hz. The probability that Poissonian noise produces by chance such a power in the $550 - 555$~Hz frequency range is 3.7$\times10^{-4}$.
The large overlap between the $N_{PDS}=120$ overlapping time intervals considered in our search prevented us from considering them statistically independent. To assess the significance of the observed power value and to ensure that rapid variations in the burst light curve did not affect the distribution of noise powers, we relied on simulations based on the procedure described by \cite{Bilous2019} (see also \citealt{Bult2021,Li2022}). We simulated $10^6$ thermonuclear burst light curves drawing photon arrival times from a Poisson distribution. 
To test that the simulated light curves correctly described the counting noise, we checked that the distribution of 100$\times$7920 noise powers\footnote{We generated 100 light curves, each covering an 8\,s time interval in the burst rise where BOs were detected, and subsequently computed the PSDs with a frequency resolution of 1/8\,Hz. This resulted in 7920 frequencies in the 10--1000 \,Hz range. We did not consider frequencies below 10 Hz to avoid contributions from red noise.} in the 10--1000\,Hz frequency range had an average value of 1.9978(14) at 1$\sigma$ confidence level, compatible with the expected value from a $\chi^2$ distribution with two degrees of freedom. For the real data, the average value (in the same frequency range) was 1.9970(23).

We used the acceptance-rejection method \citep{vonNeuman} to generate simulated light curves (LC in the following) with a count rate closely resembling that observed. To associate an average observed count rate $LC(t)$ to each 1-s bin, we first modeled the LC observed by {\it HST} fitting it with a spline. Then, we generated $LC_{\text{max}} \times N_{\text{bins}}$ pairs of uniformly distributed random variables $(\ell, \tau)$ in the range $[0, LC_{max}]$ and $[0,N_{bin}]$, respectively. Here, $LC_{\text{max}}$ is the maximum observed count rate and $N_{\text{bins}}=127$ is the number of 1 s-long bins that cover the considered interval. We then removed from the simulated event list the pairs $(\ell, \tau)$ with $\ell > LC(t)$. In each of the simulated datasets, we repeated the coherent search strategy in the 550-555 Hz frequency range obtaining $p=3.7\times10^{-2}$ of the times a signal with a power equal to or higher than\footnote{For comparison, if the $N_{PDS}$ spectra were considered independent, the probability of drawing a signal with a power of $P_{max}$ would have been $4.5\times10^{-2}$.} $P_{max}$. We consider such a value 
a robust estimate of the probability that a signal with a power equal to or higher than the maximum observed is produced by noise. 
A similar signal is expected to appear by chance once in every $\sim 27$ FUV bursts observed, and to our knowledge this is the first UV thermonuclear burst and only ever detected by STIS-HST. 
In addition, the false alarm rate would be reduced to $3 \times 10^{-3}$ if we restricted our search to the burst rise only, i.e. the stage where X-ray BOs have only been detected from \EXO. However, given the low significance, the signal observed in the power spectrum could be a statistical fluctuation. Future observations of this or other sources in the UV and/or optical bands may confirm or rule out the presence of BOs at energies lower than the X-ray band.

\begin{figure}
   \centering
   \resizebox{\hsize}{!}{\includegraphics{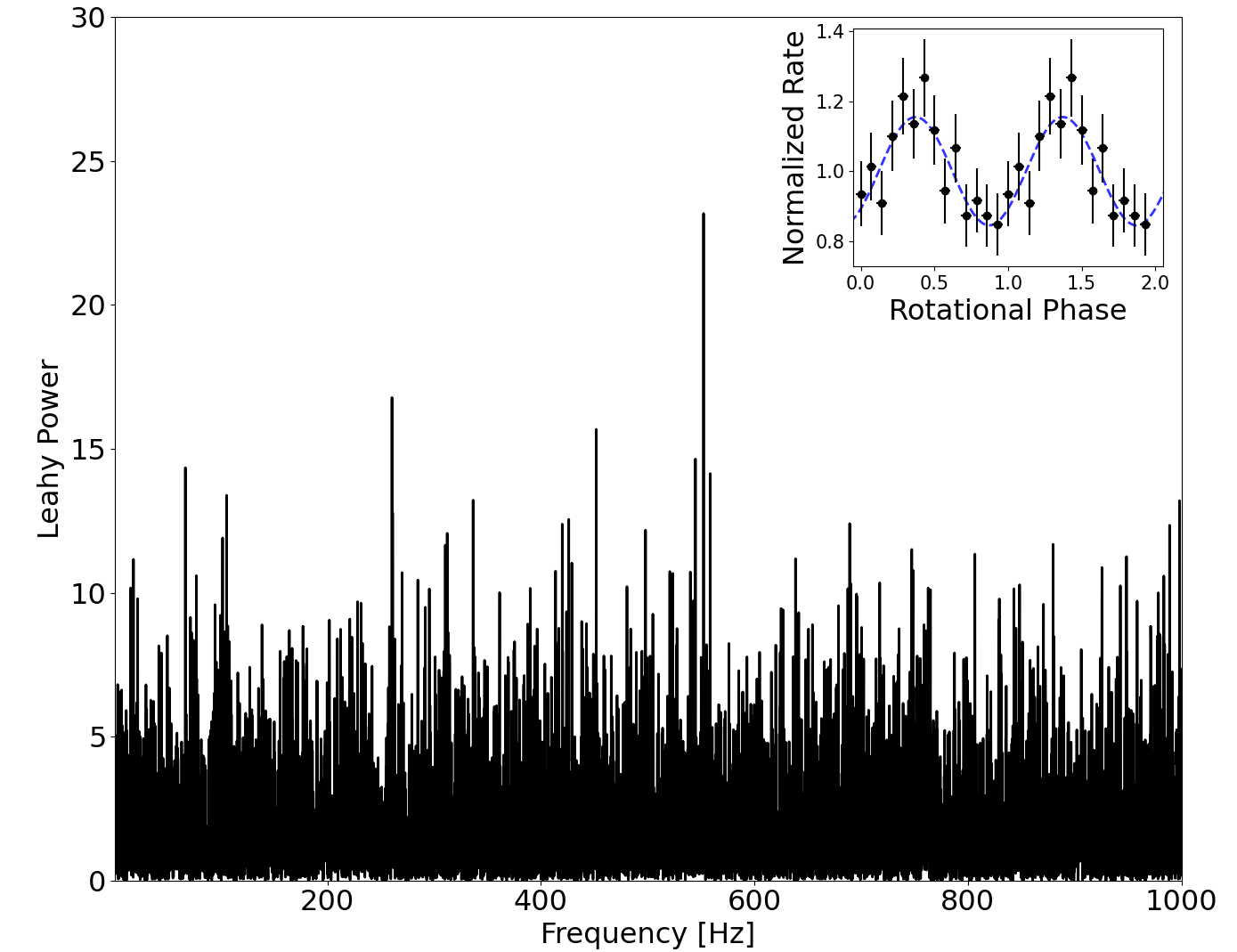}}
   \centering
      \caption{Fourier power density spectrum from the FUV (120--160\,nm) light curve collected with STIS on board {\it HST} during an 8-s observation chunk in the interval of (5--13)\,s since $T_{ref} = 52689.0772718$\,MJD during the rise of the Type I burst. 
      The light curve was rebinned to 250\,\textmu s, yielding a Nyquist frequency of 2\,kHz. 
      The strong peak is at a frequency of $\sim$ 552.392\,Hz. In the inset plot, we show the normalized pulse profile obtained by folding the light curve in the same time interval adopted to extract the power density spectrum using 14 phase bins per period cycle. The time series is folded at the frequency of 552.392\,Hz. The blue dashed line shows the best-fit decomposition with one harmonic. The background-subtracted pulsed fraction is (16.1$\pm$5.4)\% at 1$\sigma$ confidence level}. Two cycles are plotted for clarity.
         \label{PS}
   \end{figure}

The STIS background count rate was estimated by selecting photons in the slit channels 400–900 and 1100–1800 to avoid the source contribution, within the same time interval as the BO candidate. We then normalized the average count rate to the total number of source slit channels (15). The total count rate (source plus background) was 195\,\(\pm\)\,14\,counts\,s\(^{-1}\), with an estimated background contribution of approximately 3\% (BKG\(_{\mathrm{HST}}\) = 5\,\(\pm\)\,2\,counts\,s\(^{-1}\), normalized to 15 slit channels). The background-subtracted pulsed fraction 
(the semi-amplitude of the modulation divided by the average source count rate) of the FUV oscillations we tentatively identified during the burst rise is $\sim 16\%$ (see inset plot in Fig.\,\ref{PS}). 
The FUV average pulse profile in the 120--160\,nm band has an almost sinusoidal shape (see inset plot in Fig.\,\ref{PS}) similar to that reported in the X-ray band \citep{Galloway2010}. 

X-ray BOs were searched throughout the entire burst observed simultaneously by {\it RXTE}, but they were not detected. The upper limit on the pulse amplitude was estimated by computing the PDS using ToAs in the 3-20\,keV energy band and in the time interval (5--13)\,s since $T_{ref} = 52689.0772718$\,MJD
and following the procedure reported by \cite{Vaughan1994}. We derived a 3$\sigma$ upper limit of $\sim$ 7\%.  The analysis performed by \cite{Galloway2010} and \cite{Bilous2019} did not find BOs during this burst, either.

\subsection{FUV coherent periodicity search}
\label{blind search}

We also searched for coherent persistent pulsations in \EXO\; using all data in the FUV band acquired with {\it HST}-STIS. The total exposure time was 7.5\,hr, covering about 2 orbits of the binary system. We performed a blind periodicity search in a restricted two-dimensional parameters space starting from the ephemeris reported in \cite{Wolff2009} which was derived from the X-ray orbital eclipses. We demodulated the ToAs by correcting them
for the orbital R{\o}mer delay in a circular orbit, which is given by $\Delta_{RB} =x_p \ \sin{M_{\rm an}}$, where $x_p$ represents the projected semi-major axis of the pulsar orbit and $M_{\rm an}$ denotes the mean anomaly. For a circular orbit, the mean anomaly simplifies to $M_{\rm an}\equiv\Omega_b$($t$-T$_{\rm asc}$), where $\Omega_b\equiv 2\pi/P_{\rm orb}$ is the angular velocity, $P_{\rm orb}$ is the orbital period, and T$_{\rm asc}$ denotes the epoch of passage at the ascending node. 
We varied $T_{\rm asc}$ within the range 52689.1335\,MJD$<T_{\rm asc}<$52689.1393\,MJD with a step size of 10$^{-5}$\,MJD and the projection of the pulsar semi-major axis $x_p$ within the range 0.05 lt-s $< x_p <$1 lt-s (light-seconds) with a step of $2.5 \times 10^{-3}$\,lt-s. The orbital period was kept fixed at $P_{\rm orb}$\,=\,0.1593377\,d \citep{Wolff2009}. 
The starting value of $T_{\rm asc}$ was estimated by subtracting $P_{\rm orb}/4$ from the X-ray mid-eclipse epoch measured in the simultaneous {\it RXTE} observations.
The search interval for the pulsar spin frequency was limited within the frequency range 545\,Hz\,$ < \nu < $565\,Hz.
We found no evidence for FUV pulsations and put an upper limit on the FUV pulsed sinusoidal amplitude below 1.7\% at 3$\sigma$ confidence level.

\begin{figure}
   \centering
   \resizebox{\hsize}{!}{\includegraphics{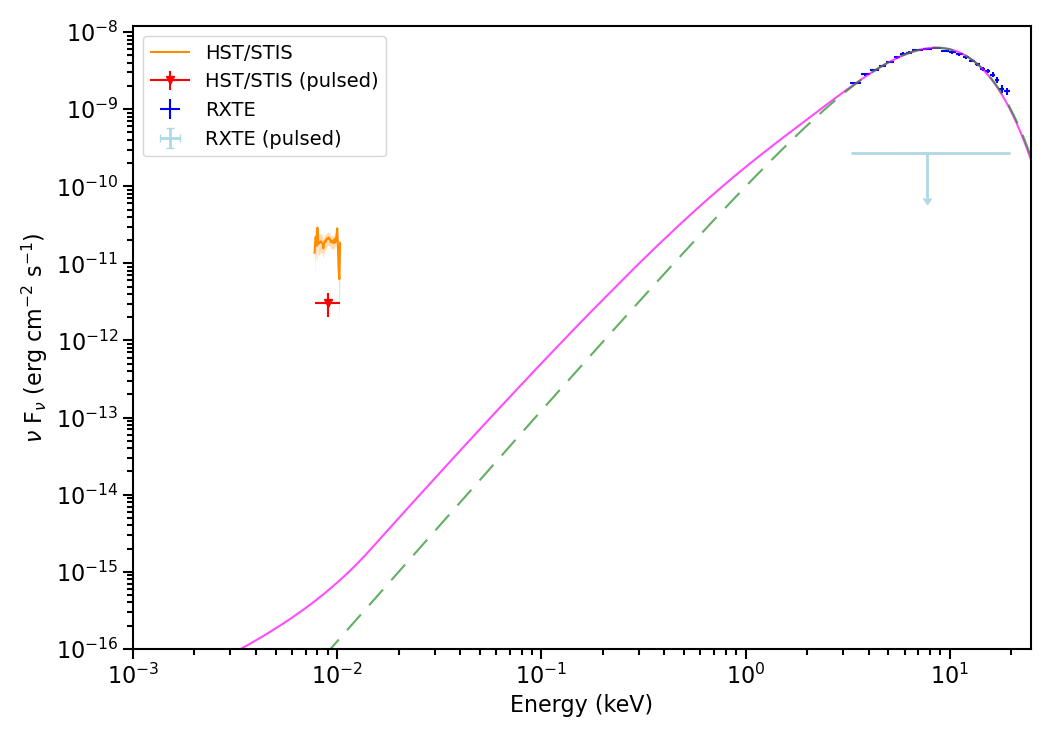}}
   \centering
      \caption{Spectral energy distribution for the total and pulsed emission of \EXO\;corrected for the interstellar extinction. The X-ray spectrum is extracted during the thermonuclear burst in the time interval (5–13) s since $T_{ref}$ (TDB), while the FUV spectrum is extracted during the first 80\,s of the burst. The X-ray spectrum acquired with {\it RXTE} in the 3--20\,keV energy band is plotted in blue. The X-ray upper limit on pulsed emission is marked by a light blue arrow. The FUV fluxes measured with {\it HST} are shown in orange.
      The FUV pulsed flux with 
 the corresponding error bars is plotted in red. The magenta line represents the fit to the X-ray spectrum using the \texttt{burstatmos} model \citep{Suleimanov}, corrected for interstellar extinction and extrapolated down to the FUV band. The green dashed line shows the fit using the \texttt{bbodyrad} model in XSPEC, also corrected for interstellar extinction and extrapolated to the FUV range. The blackbody has a temperature of $\sim 2.2$\,keV and an emitting radius of $\sim 4$\,km (assuming a source distance of 7.1\,kpc; see text).}
         \label{SED}
   \end{figure}

\section{Discussion}
\label{Discussion}

Our study reveals a candidate detection of FUV burst oscillations at the NS spin period of $\sim$1.8\,ms, observed during an 8\,s interval of a thermonuclear burst in the binary MSP \EXO. We observed the candidate signal during the rise of the FUV emission, marking the first such detection in any Type I bursting source. The candidate signal suggests a pulse amplitude of (16.1$\pm$5.4)\% in the 120--160\,nm FUV band, which is compatible within the uncertainties with the X-ray BO amplitude observed on other occasions from the same source ($\sim 21\%$, \citealt{Galloway2010}), as well as similar to other LMXBs \citep{Galloway2020, Bhattacharyya2022}.
For a source distance of $\sim$ 7.1\,kpc \citep{Galloway2008a, Galloway2008b}, 
such an amplitude would imply a FUV pulsed luminosity of  $L_{\rm FUV, pulsed} = 0.16\,L_{\rm FUV} \sim 8 \times 10^{33}$\,erg\,s$^{-1}$ in the 120$-$160\,nm band.
The FUV burst light curve lagged the X-ray burst light curve by about 4\,s \citep{Hynes2006, Knight2025}, a delay that is consistent with the FUV emission originating from reprocessed X-rays on the surface of the companion star and 
the outer accretion disk regions.
This may suggest that the FUV BOs could also possibly originate from the reprocessing of X-ray BOs. However, such an interpretation can be ruled out based on different arguments. Firstly, X-ray BOs were not detected in the burst, with a 7\% upper limit on the pulse amplitude. More crucially, reprocessing should retain coherence over timescales of about half the FUV BOs period (or $\sim 1$ ms) not to be smeared out. 
For an orbital separation of $\sim\,$few$\times 10^{11}$cm and companion star radius of $\sim 2 \times 10^{10}$cm the time of flight difference of reprocessed FUV photons coming from different regions of the companion's atmosphere would amount to $\sim 100-200$\,ms, far too long to maintain coherence.
Moreover, typical reprocessing times from X-ray photons to optical-UV photons inside the companion's magnetosphere are of order $0.3-1$s \citep{Basko1973, Basko1974, Cominsky1987}. 
We conclude that FUV BOs cannot be due to reprocessing of X-ray BOs. 

The possibility that FUV oscillations originate from blackbody-like emission at the NS surface can be ruled out as well. 
We compared the spectral energy distribution of the total and candidate pulsed emission in the X-ray and FUV bands. First, we fitted the X-ray spectrum using the \texttt{bbodyrad} model in \textsc{XSPEC}. The blackbody temperature at the peak of the X-ray burst was $(2.18 \pm 0.03)$\,keV, and the inferred radius of the emitting region was $(4.2 \pm 0.1)$\,km. 
Extrapolating this model to the FUV band yields a candidate pulsed flux more than four orders of magnitude lower than the observed value (see Fig.\,\ref{SED}.)
We then applied the NS atmosphere model \texttt{burstatmos} (\citealt{Suleimanov} and references therein), fixing the hydrogen mass fraction to 1 (pure H atmosphere), the NS mass to $1.4\,\mathrm{M}_\odot$, the radius to 10\,km, and the distance to 7.1\,kpc. In Fig.~\ref{SED}, the magenta line shows the \texttt{burstatmos}\footnote{\url{https://github.com/jmjkuu/burstatmos?tab=readme-ov-file}} model fitted to the X-ray spectrum after fixing the hydrogen column density to $7 \times 10^{20}\,\mathrm{cm}^{-2}$ \citep{Degenaar2011}, and fitting the ratio $g_{\mathrm{rad}}/g$, where $g_{\mathrm{rad}}$ is the radiative acceleration (as defined in \citealt{Suleimanov2012} for non-isotropic Compton scattering) and $g$ is the NS surface gravity. The best-fit value obtained is $g_{\mathrm{rad}}/g$=$0.62(1)$. However, also in this case, the extrapolated model underpredicts the FUV pulsed flux by several orders of magnitude.

Conversely, by solving for the brightness temperature required to emit a luminosity of $\sim 8 \times 10^{33}$\,erg\,s$^{-1}$ from the surface of a NS with a 15 km radius, we derive a value of $\sim 10^{11}$\,K.   
Of course, the X-ray spectrum of the source is $\sim 4$ orders of magnitude fainter than this temperature would imply. A lower limit on the brightness temperature of $\sim 10^{8}$\,K, is obtained by considering a spherically emitting region with light crossing time of  $\sim 1$\,ms.

Brightness temperatures reaching $\sim 10^{11}$\,K approach the typical threshold for the onset of coherent emission \citep{Cordes2004, Melrose2017}, which might suggest that the candidate FUV BOs could arise from an as-yet unidentified coherent process. Alternatively, they might be produced by beamed synchrotron radiation from heated charged particles near the light cylinder radius. However, if the presence of FUV BOs is confirmed, their origin would remain unexplained within the framework of current emission models.



UV millisecond pulsations represent a recently discovered phenomenon in binary MSPs, and their interpretation remains under debate. To date, they have been firmly confirmed in only two systems. The transitional MSP PSR\,J1023+0038 shows pulsed emission in the 165–310\,nm band, with a luminosity of $\sim 10^{31}$\,erg\,s$^{-1}$ \citep{Miraval2022}, while the accreting MSP SAX\,J1808.4--3658 exhibits a pulsed luminosity of $\sim 10^{32}$\,erg\,s$^{-1}$ in the same band \citep{Ambrosino2021}. These values correspond to brightness temperatures of $\sim 10^9$ and $\sim 10^{10}$\,K, respectively, assuming the UV/optical pulsations originate at the NS surface. In the case of SAX J1808.4--3658, the optical/UV pulses are not generated during thermonuclear bursts.
If the presence of UV pulsations in EXO\,0748–676 is confirmed, its even higher pulsed luminosity would pose a significant challenge to current emission models.



\section{Conclusions}
\label{Conclusion}
We analyzed simultaneous observations of the NS-LMXB \EXO\ during a thermonuclear burst detected in 2003 by \textit{HST} and \textit{RXTE}. In our search for BOs in the FUV and X-ray bands, we found possible indication for a sinusoidal FUV signal at a frequency of $\sim$552\,Hz, potentially the first such detection in this band from any LMXB.
Interpreting the properties of this signal within the framework of current models for optical/UV emission from NS-LMXBs is not straightforward and raises important open questions. What physical mechanism can produce millisecond pulsations in the UV during thermonuclear bursts? What conditions are required for such emission to occur?
Future investigations will aim to address these questions by carrying out similar searches in this and other systems, and by extending the search to the optical band, including sources with unknown NS spin periods.
 

\begin{acknowledgements}
We thank the referee for very useful comments and suggestions. This paper is based on observations with the NASA/ESA Hubble Space Telescope, obtained at the Space Telescope Science Institute, which is operated by AURA, Inc., under NASA contract NAS5-26555. 
FCZ is supported by a Ram\'on y Cajal fellowship (grant agreement RYC2021-030888-I). FA, GI, AP, LS, and DdM acknowledge financial support from the Italian Space
Agency (ASI) and National Institute for Astrophysics (INAF)
under agreements ASI-INAF I/037/12/0 and ASI-INAF
n.2017-14-H.0, from INAF Research Grant `Uncovering the
optical beat of the fastest magnetised neutron stars (FANS)'. FA, GI, AP, and LS also acknowledge funding
from the Italian Ministry of University and Research (MUR),
PRIN 2020 (prot. 2020BRP57Z) `Gravitational and Electromagnetic-wave Sources in the Universe with current and next generation detectors (GEMS)'.
This work was also partially supported by the program Unidad de Excelencia Maria de Maeztu CEX2020-001058-M. GB acknowledges support from ASI/INFN grant n. 2021-43-HH.0. AP acknowledges support from the Fondazione Cariplo/Cassa Depositi e Prestiti, grant no. 2023-2560.\\
\textit{Facilities:} ADS, HEASARC, {\it HST}, {\it RXTE}\\
\textit{Software:} HEASoft (v6.31), IRAF (v2.16), python3 (v3.10.8), \texttt{stis$\_$photons}

\end{acknowledgements}

\bibliography{biblio}{}
\bibliographystyle{aa}

\end{document}